\documentclass[journal]{IEEEtran}
\IEEEoverridecommandlockouts

\usepackage{subcaption}
\usepackage[dvipsnames]{xcolor}
\usepackage{siunitx}
\usepackage{xspace}
\usepackage{comment}
\usepackage{array}
\usepackage{hyperref}
\usepackage{natbib}

\usepackage{cite}
\usepackage{amsmath,amssymb,amsfonts}
\usepackage{algorithmic}
\usepackage{graphicx}
\usepackage{textcomp}
\usepackage{xcolor}
\usepackage{url}
\usepackage{booktabs}
\usepackage{makecell}
\usepackage[most]{tcolorbox}
\usepackage{tabularx}
\usepackage{booktabs}
\usepackage{multirow}
\usepackage{booktabs}
\usepackage{setspace}

\def\BibTeX{{\rm B\kern-.05em{\sc i\kern-.025em b}\kern-.08em
    T\kern-.1667em\lower.7ex\hbox{E}\kern-.125emX}}

\begin{document}

\doublespacing

\title{\LARGE{Acoustic and Attentional Challenges Shape the Balance Between Attended and Competing Speech Tracking Differently}
 
\thanks{This work was supported by the ELLIIT Strategic Research Programme. We thank all participants for taking part in this study and the research audiologists at Eriksholm Research Centre for their assistance with data collection.}
}

\author{\IEEEauthorblockN{%
    Johanna Wilroth\IEEEauthorrefmark{1}\IEEEauthorrefmark{6},
    Martin A. Skoglund\IEEEauthorrefmark{1}\IEEEauthorrefmark{3}}, Emina Alickovic\IEEEauthorrefmark{1}\IEEEauthorrefmark{3}, and Martin Enqvist\IEEEauthorrefmark{1} \\
    \vspace{0.8cm}
  \IEEEauthorblockA{\IEEEauthorrefmark{1}%
  \textit{Dept. of Electrical Engineering}
\textit{Linköping University}
Sweden. %
    \texttt{\{firstname.lastname\}@liu.se}
}\\

  \IEEEauthorblockA{\IEEEauthorrefmark{3}%
  \textit{Eriksholm Research Centre, Oticon A/S} Denmark. \texttt{\{mnsk,eali\}@eriksholm.com}} \\
  \IEEEauthorblockA{%
  \thanks{\IEEEauthorrefmark{6} Author to whom any correspondence should be addressed.}}
   
}
    
\onecolumn
\maketitle

\def\tvpersonleft#1#2#3{
\begin{scope}[shift={#1}, rotate=#2]
    \clip (-1,-0.6) rectangle (1,0.8);
    \node[#3,minimum size=1.2cm,rotate=#2]{};
    \draw (-1,-0.6) rectangle (1,0.8);
\end{scope}
}

\def\tvpersonright#1#2#3{
\begin{scope}[shift={#1}, rotate=#2]
    \clip (-1,-0.6) rectangle (1,0.8);
    \node[#3,mirrored,minimum size=1.2cm,rotate=#2]{};
    \draw (-1,-0.6) rectangle (1,0.8);
\end{scope}
}

\def\speaker#1#2#3#4{
\begin{scope}[shift={#1}, rotate=#2]
    \draw[#4,thick] (0*#3,0.5*#3)--(0.5*#3,1*#3)--(0.5*#3,-1*#3)--(0*#3,-0.5*#3)--(0*#3,0.5*#3)--(-0.25*#3,0.5*#3)--(-0.25*#3,-0.5*#3)--(0*#3,-0.5*#3);
\end{scope}
}

\definecolor{pyblue}{rgb}{0.121, 0.466, 0.705}
\definecolor{pyorange}{rgb}{1.000, 0.498, 0.055}
\definecolor{pygreen}{rgb}{0.173, 0.627, 0.173}
\definecolor{pyred}{rgb}{0.839, 0.153, 0.157}
\definecolor{johanna}{RGB}{150,50,200}
\newcommand{\johanna}[1]{\textcolor{johanna}{\textbf{#1}}}

\makeatletter

\let\otoprule\toprule

\newenvironment{notationtabular@modern}[3]{%
  \addvspace{1\baselineskip}
  \noindent\textbf{#1}\par\vspace{0.3em}
  \noindent\centering
  \begin{tabular}{@{} c l @{}}
    \toprule
    \textbf{#2} & \textbf{#3} \\
    \midrule
}{%
    \bottomrule
  \end{tabular}\par
}

\newenvironment{notationtabular@old}[3]{%
  \addvspace{1\baselineskip}
  \begin{tabular}{>{\centering}m{0.18\textwidth}|p{0.67\textwidth}}
    \multicolumn{2}{l}{\rlap{\hspace*{-3ex}\raisebox{1ex}{\textbf{#1}}}}\\
    \hline
    \textbf{#2} & \parbox{\linewidth}{\centering\textbf{#3}} \\
    \hline
    \hline
}{%
    \hline
  \end{tabular}
}

\newenvironment{notationtabular}
  {\notationtabular@modern}
  {\endnotationtabular@modern}

\makeatother

\let\textabbr\textsc
\newcommand{\abbrERP}{\texorpdfstring{\textabbr{erp}}{ERP}\xspace}
\newcommand{\abbrTRF}{\texorpdfstring{\textabbr{trf}}{TRF}\xspace}
\newcommand{\abbrHI}{\texorpdfstring{\textabbr{hi}}{HI}\xspace}
\newcommand{\abbrNH}{\texorpdfstring{\textabbr{nh}}{NH}\xspace}
\newcommand{\abbrEEG}{\texorpdfstring{\textabbr{eeg}}{EEG}\xspace}
\newcommand{\abbrMEG}{\texorpdfstring{\textabbr{meg}}{MEG}\xspace}
\newcommand{\abbrAAD}{\texorpdfstring{\textabbr{aad}}{AAD}\xspace}
\newcommand{\abbrDTU}{\texorpdfstring{\textabbr{dtu}}{DTU}\xspace}
\newcommand{\abbrERH}{\texorpdfstring{\textabbr{erh}}{ERH}\xspace}
\newcommand{\abbrICA}{\texorpdfstring{\textabbr{ica}}{ICA}\xspace}
\newcommand{\abbrSNR}{\texorpdfstring{\textabbr{snr}}{SNR}\xspace}
\newcommand{\abbrNR}{\texorpdfstring{\textabbr{nr}}{NR}\xspace}
\newcommand{\abbrHA}{\texorpdfstring{\textabbr{ha}}{HA}\xspace}
\newcommand{\abbrFIR}{\texorpdfstring{\textabbr{fir}}{FIR}\xspace}
\newcommand{\abbrBBN}{\texorpdfstring{\textabbr{bbn}}{BBN}\xspace}
\newcommand{\abbrFF}{\texorpdfstring{\textabbr{ff}}{FF}\xspace}
\newcommand{\abbrAV}{\texorpdfstring{\textabbr{av}}{AV}\xspace}
\newcommand{\abbrITD}{\texorpdfstring{\textabbr{itd}}{ITD}\xspace}
\newcommand{\abbrILD}{\texorpdfstring{\textabbr{ild}}{ILD}\xspace}
\newcommand{\abbrELU}{\texorpdfstring{\textabbr{elu}}{ELU}\xspace}
\newcommand{\abbrAESPA}{\texorpdfstring{\textabbr{aespa}}{AESPA}\xspace}

\newcommand{\reals}{\mathbb{R}}
\newcommand{\complexes}{\mathbb{C}}
\newcommand{\rationals}{\mathbb{Q}}
\newcommand{\naturals}{\mathbb{N}}
\newcommand{\integers}{\mathbb{Z}}

\newcommand{\textfilename}[1]{\mbox{\emph{#1}}}
\newcommand{\styname}[1]{\texorpdfstring{{\fontsize{0.9em}{1\baselineskip}\selectfont\mbox{\textsf{#1}}}}{#1}}
\newcommand{\classoption}[1]{\mbox{\texttt{#1}}}
\newcommand{\texcommand}[1]{\mbox{\texttt{\textbackslash #1}}}
\newcommand{\envname}[1]{\mbox{\texttt{#1}}}

\newcommand{\rtthesis}{\emph{\styname{rtthesis}}\xspace}

\newcommand{\usc}{\hspace{0.1ex}\rule{1ex}{0.1ex}\hspace{0.1ex}}

\begin{abstract}
Selective attention enhances cortical tracking of attended speech, but the role of ignored speech remains less clear. In particular, it is unknown whether attended and ignored speech contribute to cortical tracking in a fixed balance or whether this balance changes with listening conditions. We addressed this question across six electroencephalography (\abbrEEG) datasets spanning acoustic and attentional challenges, hearing impairment, and hearing-aid processing. We systematically varied the relative weighting of attended and competing speech and studied how their combined contribution affected cortical speech tracking.
Across most listening conditions, cortical tracking was improved when we included competing speech, compared with attended speech alone. Importantly, the balance between attended and competing speech changed across conditions. Increasing acoustic difficulty increased both the benefit and the optimal contribution of competing speech, with the same general pattern observed in normal-hearing and hearing-impaired listeners. In the hearing-aid datasets examining noise-reduction conditions, the contribution of competing speech was more limited. Attentional challenges also changed the contribution of competing speech, but differently from acoustic challenges.
These findings show that attended and competing speech do not contribute to cortical tracking in a fixed balance. Rather, their balance depends on the listening challenge. Although these results do not define a single optimal ratio, they provide a step toward hearing devices that adapt not only to the attended talker, but also to how competing speech is balanced in complex listening environments.
\end{abstract}

\begin{IEEEkeywords}
Auditory attention decoding, speech mixtures, EEG, competing speech attenuation, acoustic and attentional challenges

\end{IEEEkeywords}

\section{Introduction}
\label{sec:int}

Understanding one talker among competing speakers is a fundamental challenge for the auditory system. The classic cocktail party problem introduced by \citet{1cherry1953some} demonstrated that listeners can selectively follow one speech stream while retaining little of the linguistic content of an ignored stream. Yet ignored speech is not completely excluded: listeners remain sensitive to changes in its physical characteristics, and important information such as their own name can occasionally attract attention \citep{2moray1959}. These observations motivated different explanations of selective attention. Whereas \citet{3broadbent1963} proposed that unattended information is filtered at an early stage of processing, \citet{4TREISMAN1964} argued that ignored information is attenuated rather than completely removed.

The extent to which competing information is processed may depend on the listening conditions. Perceptual load theory proposes that distractor processing depends on the demands placed on perceptual resources, with greater processing of irrelevant information when spare capacity is available \citep{5lavie1995}. In complex auditory scenes, successful listening also depends on the formation of distinct auditory objects and the selective allocation of attention to the relevant object \citep{6SHINNCUNNINGHAM2008}. Acoustic degradation introduces an additional challenge. When the incoming speech signal does not readily match stored linguistic information, greater reliance on working memory and explicit cognitive processing may be required, as proposed by the ease of language understanding (\abbrELU) model \citep{12ronnberg2013}. Together, these accounts suggest that the processing of competing speech is not fixed, but may depend on both the acoustic and attentional challenges of listening.

Cortical speech tracking provides a means of investigating how these challenges affect the processing of competing speech streams. Regression-based approaches relate ongoing \abbrEEG{} activity to continuous features of speech, initially introduced as auditory evoked spread spectrum analysis (\abbrAESPA) \citep{7lalor2010} and subsequently widely applied using temporal response functions ({\abbrTRF}s). During selective listening, cortical activity tracks both attended and ignored speech, with stronger tracking generally observed for the attended stream \citep{8DingSimon2012a}. This attentional modulation can be detected from single-trial \abbrEEG{} and used to identify the attended talker \citep{13o2015attentional}. Cortical speech tracking also changes with the listening conditions, including reverberation, background noise, spatial separation, and speech intelligibility \citep{14Fuglsang2017,15das2018,Vanthornhout2018,alickovic2020front, ELLIITdataset}. Thus, cortical speech tracking provides access to both speech streams and to changes associated with the listening environment.

What remains less understood is the balance between attended and competing speech. Most studies quantify attended- and ignored-speech tracking separately. Such estimates establish attentional selectivity but do not determine the relative weighting of attended and competing speech when both streams are considered jointly. This gap is relevant across distinct listening manipulations: acoustic challenges alter the sensory conditions supporting stream segregation, whereas attentional challenges alter the demands associated with selecting, maintaining, or switching the task-relevant stream. It therefore remains unclear whether cortical speech tracking is best characterised by attended speech alone or by a weighted combination of attended and competing speech, and whether this balance changes across acoustic and attentional conditions. Characterising the attended-competing balance would clarify the extent to which competing speech remains reflected in cortical tracking while attention is directed towards another talker. 

Hearing impairment may further affect this balance. Reduced access to acoustic cues may limit the segregation of competing sound sources, while hearing aid (\abbrHA) processing can modify the signals available to the listener. Previous work has shown that \abbrHA processing can change cortical tracking of attended speech, competing speech, and background noise \citep{alickovic2020front, ELLIITdataset, Andersen2021, WilrothENEURO, mai2026hearing}. In this study, we examine complementary aspects of this question. Acoustic challenge is varied across signal-to-noise ratios ({\abbrSNR}s) in hearing-impaired listeners using standard {\abbrHA}s. Other conditions maintain a fixed acoustic {\abbrSNR} while varying \abbrHA configuration. These conditions allow us to determine whether acoustic-condition dependence of the attended-competing balance is also observed in hearing-impaired listeners and whether the balance differs across \abbrHA processing conditions. They do not represent a matched comparison between normal-hearing and hearing-impaired listeners or a complete comparison of all factors of \abbrHA configurations and noise reduction ({\abbrNR}) settings.

Understanding this balance may also be relevant to future neuro-steered hearing devices. Identifying the attended talker does not determine how strongly competing speech should be attenuated. Competing talkers may interfere with the target, but they may also contain information that the listener needs to monitor or attend subsequently. Future hearing devices may therefore need to determine not only which talker to enhance, but also how competing speech should be treated as the listening conditions change. 

\citet{Wilroth2026IFAC} introduced a weighted-mixture approach in which the relative contributions of attended and ignored speech were varied systematically. In a single dataset, cortical speech tracking followed an inverted-U weighting function, with the strongest tracking occurring when both speech streams contributed. This study extends this approach across multiple independent \abbrEEG{} datasets spanning acoustic and attentional conditions, normal-hearing and hearing-impaired listeners, and \abbrHA processing conditions. We systematically varied the relative weighting of attended and competing speech to determine whether the weighting associated with strongest cortical speech tracking changes across listening conditions. The estimated weighting is a model-derived descriptor of the attended-competing balance and should not be interpreted as a literal percentage or direct neural proportion of competing speech tracked by the brain.

We tested three hypotheses and addressed one exploratory question (EQ): 
\begin{itemize}
    \item[] \textbf{H1: Competing-speech weighting}
    \begin{itemize}
        \item[] Cortical speech tracking will be maximal at a non-zero competing-speech weight, with the optimal relative weighting of attended and competing speech adapting to the listening conditions.
    \end{itemize}
    \item[] \textbf{H2: Acoustic versus attentional challenges}
    \begin{itemize}
        \item[] The attended-competing balance will depend on the source of the listening challenge, with acoustic and attentional challenges producing different changes in the contribution of competing speech.
    \end{itemize}
    \item[] \textbf{H3: Generalisation to hearing-impaired listeners}
    \begin{itemize}
        \item[] Acoustic-condition dependence of the attended-competing balance will also be present in hearing-impaired listeners using {\abbrHA}s.
    \end{itemize}
    \item[] \textbf{EQ1: Hearing-aid processing}
    \begin{itemize}
        \item[] At a fixed acoustic {\abbrSNR}, does the attended-competing balance vary across \abbrHA processing conditions, including \abbrNR settings? 
    \end{itemize}
\end{itemize}

\newpage
\section{Experimental Datasets}
\label{sec:exp}

In this study, we analyzed six \abbrEEG{} datasets designed to capture a wide range of real-world listening challenges. These datasets included bottom-up acoustic manipulations, such as varying the \abbrSNR{} between the target speaker and background babble noise ({\abbrBBN}), as well as top-down cognitive manipulations involving sustained attention, attentional switching, and audiovisual conversational tasks. They further comprised recordings from both normal-hearing and hearing-impaired participants, the latter using {\abbrHA}s with either {\abbrNR} enabled or disabled.

All studies were conducted in accordance with the Declaration of Helsinki, and written informed consent was obtained from all participants prior to participation.

\begin{figure}[t]
    \centering
    \includegraphics[width=0.95\linewidth]{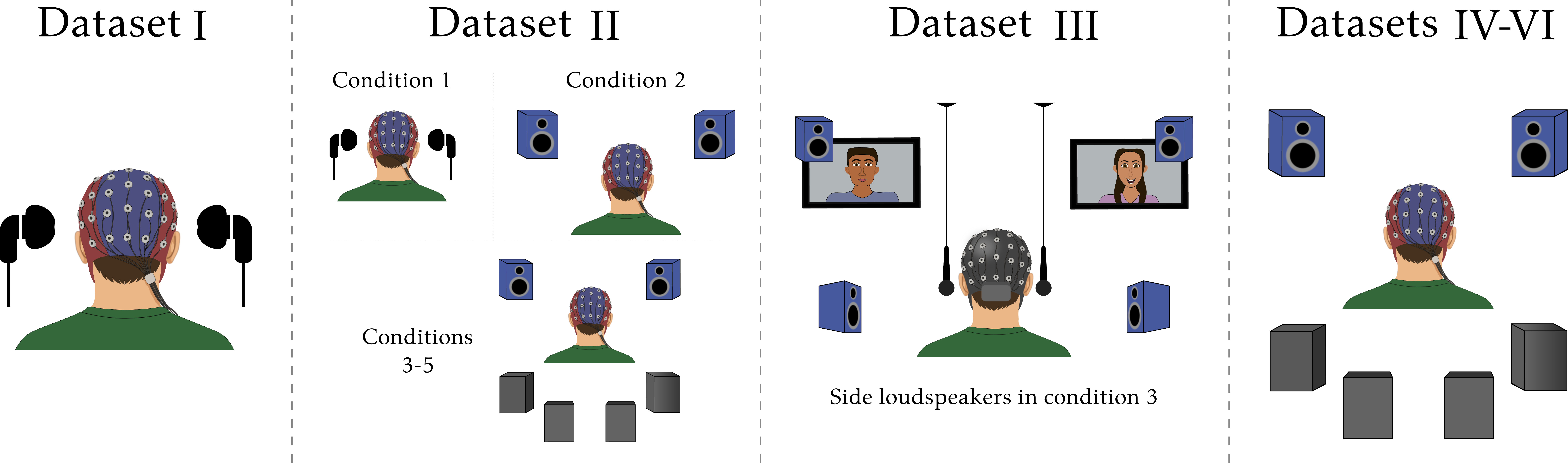}
    \caption{Illustration of the experimental paradigms used in
    datasets I--VI.}
    \label{fig:figure1}
\end{figure}

\begin{table}[t]
\centering
\footnotesize

\caption{\textbf{Overview of the datasets and experimental conditions.}
\abbrFF{}: free-field;
\abbrAV{}: audiovisual;
\abbrBBN{}: background babble noise;
\abbrHA{}: hearing aid;
\abbrNR{}: noise reduction.}
\label{tab:dataset-comparison}

\textbf{Dataset characteristics}

\vspace{0.5em}
\small
\renewcommand{\arraystretch}{1.5}

\begin{tabular}{@{}lcccccc@{}}
\toprule
 & \textbf{I} & \textbf{II} & \textbf{III} &
   \textbf{IV} & \textbf{V} & \textbf{VI} \\
\midrule
Participants
    & 18 & 23 & 21 & 31 & 35 & 17 \\
Hearing
    & \abbrNH{} & \abbrNH{} & \abbrNH{} &
      \abbrHI{} & \abbrHI{} & \abbrHI{} \\
Hearing aids
    & No & No & No & Yes & Yes & Yes \\
Age group
    & Young--middle & Young--middle & Young--middle & Middle--older & Middle--older & Middle--older \\
\abbrBBN{}
    & No & Yes/No & No & Yes & Yes & Yes \\
Presentation
    & Insert & Insert/\abbrFF{} & \abbrAV{} \abbrFF{} &
      \abbrFF{} & \abbrFF{} & \abbrFF{} \\
Conditions
    & 3 & 5 & 3 & 4 & 4 & 6 \\
Trials/condition
    & 20 & 32 & 9, 9, 10 & 20 & 20 & 12 \\
Trial length (\si{\second})
    & 50 & 60 & 180 & 33 & 33 & 60 \\
\abbrEEG{} system
    & BioSemi & BioSemi & mBrainTrain &
      BioSemi & BioSemi & BioSemi \\
\abbrEEG{} channels
    & 64 & 64 & 44 & 64 & 64 & 64 \\
Sampling rate (\si{\hertz})
    & 512 & 8192 & 500 & 1024 & 1024 & 8192 \\
Recorded at
    & \abbrDTU{} & \abbrERH{} & \abbrERH{} &
      \abbrERH{} & \abbrERH{} & \abbrERH{} \\
\bottomrule
\end{tabular}%

\vspace{1.5em}

\textbf{Experimental conditions}

\vspace{0.5em}

\normalsize
\renewcommand{\arraystretch}{2}

\begin{tabularx}{\linewidth}{@{}cX@{}}
\toprule
\textbf{Dataset} & \textbf{Conditions} \\
\midrule

I &
Anechoic; mild reverberation; high reverberation. \\

II &
Insert earphones without \abbrBBN{};
{\abbrFF} without \abbrBBN{};
{\abbrFF} at \abbrSNR{} \SI{-1}{\decibel},
\SI{-4}{\decibel}, and \SI{-7}{\decibel}. \\

III &
Sustained attention;
switching attention;
attending frontal \abbrAV conversation or side loudspeaker. \\

IV &
\abbrHA{}1 with \abbrNR{} enabled or disabled;
\abbrHA{}2 with \abbrNR{} enabled or disabled;
all conditions presented at \abbrSNR{} \SI{3}{\decibel}. \\

V &
\abbrHA{}1 with \abbrNR{} enabled or disabled;
\abbrHA{}3 with \abbrNR{} enabled or disabled;
all conditions presented at \abbrSNR{} \SI{3}{\decibel}. \\

VI &
\abbrSNR{} \SI{12}{\decibel},
\SI{8}{\decibel},
\SI{4}{\decibel},
\SI{2}{\decibel},
\SI{-1}{\decibel}, and
\SI{-4}{\decibel}. \\

\bottomrule
\end{tabularx}

\end{table}

\subsection{Experimental Paradigms}

The experimental paradigms are illustrated in Figure~\ref{fig:figure1}, while the characteristics of each dataset are summarized in Table~\ref{tab:dataset-comparison}. Because the datasets were collected independently by different research groups for different research questions, their preprocessing pipelines differed somewhat. Common preprocessing steps included re-referencing, band-pass and notch filtering, artifact removal using independent component analysis (\abbrICA{}), visual inspection and rejection of noisy channels, and normalization. To ensure a consistent analysis across datasets, all recordings underwent an additional standardized preprocessing pipeline before further analysis. This final processing is described in Section~\ref{sec:finalProcessing}. Below, we briefly summarize the experimental paradigm and original preprocessing for each dataset.

\subsubsection{Dataset I}

Dataset I investigated the effect of reverberation on auditory attention decoding. Participants listened to two competing audiobook stories presented dichotically through insert earphones while attending one speaker and ignoring the other. Three acoustic conditions were evaluated: anechoic, mild reverberation, and high reverberation. A 64-channel BioSemi ActiveTwo system was used to record \abbrEEG. The dataset is publicly available through Zenodo \citep{Fuglsang2018Dataset}, while the experimental design and preprocessing are described in detail by \citet{14Fuglsang2017}. Briefly, the \abbrEEG{} was re-referenced to the mastoid electrodes, band-pass and notch filtered, artifact rejection using \abbrICA{}, downsampled, and low-pass filtered to \SI{8}{\hertz}.

\subsubsection{Dataset II}

Dataset II examined the effect of acoustic scene complexity on auditory attention decoding. Participants attended one of two competing speakers presented either through insert earphones or free-field ({\abbrFF}) loudspeakers. In addition to quiet listening conditions, several conditions included multi-talker \abbrBBN presented from loudspeakers behind the participant at different \abbrSNR{} levels. A 64-channel BioSemi ActiveTwo system was used to recoded \abbrEEG{}. The experimental paradigm and preprocessing are described in \citep{alickovic2023_data2}. Similar to Dataset I, preprocessing included re-referencing, band-pass and notch filtering, artifact rejection using \abbrICA{}, removal of noisy channels, and downsampling prior to analysis.

\subsubsection{Dataset III}

Dataset III focused on cognitive aspects of selective auditory attention in an audiovisual listening environment. Twenty-four participants viewed two speakers presented on monitors positioned in front of the listener while \abbrEEG{} was recorded using a wearable 64-channel system. Three participants were excluded in this study due to bad data quality and missing trials. Three experimental conditions were analyzed: sustained attention to one speaker, attentional switching between speakers, and attending a frontal audiovisual conversation while ignoring a lateral speaker or attending the lateral speaker while ignoring the frontal conversation. The dataset was recorded using a modified Smarting Pro X amplifier with 44 scalp electrodes and 20 around-the-ear (c{\abbrEEG}rid) electrodes. Only the scalp electrodes were included in this study. Detailed descriptions of the experimental paradigm and preprocessing are provided by \citet{mobEEG}. Briefly, preprocessing consisted of re-referencing, notch and band-pass filtering, artifact removal using \abbrICA{}, downsampling, and normalization.

\subsubsection{Datasets IV--V}

Datasets IV and V investigated the influence of \abbrHA~{\abbrNR} on cortical speech tracking in older hearing-impaired listeners. Although recorded at different times using different participant groups, the two datasets followed identical experimental paradigms and are therefore analyzed together throughout this study. Participants listened to two competing frontal speakers presented through \abbrFF loudspeakers with multi-talker {\abbrBBN}. The experimental conditions differed according to \abbrHA manufacturer and whether the \abbrHA~{\abbrNR} algorithm was enabled or disabled. A 64-channel BioSemi ActiveTwo system was used to record {\abbrEEG}. The experimental paradigm and preprocessing are described in detail by \citet{WilrothENEURO}, while additional information regarding the \abbrHA processing is provided by \citep{Andersen2021,ELLIITdataset}. Briefly, preprocessing included re-referencing, band-pass and notch filtering, visual inspection, artifact removal using \abbrICA{}, and downsampling.

\subsubsection{Dataset VI}

Dataset VI examined the effect of decreasing speech intelligibility in older hearing-impaired listeners wearing standard {\abbrHA}s without {\abbrNR}. Participants attended one of two competing frontal speakers while multi-talker babble noise was presented from loudspeakers positioned behind the listener. Listening difficulty was manipulated by varying the \abbrSNR{} across six conditions. \abbrEEG{} was recorded using a 64-channel BioSemi ActiveTwo system. The experimental paradigm and preprocessing are described by \citet{payam_dataset}. Briefly, preprocessing included re-referencing, band-pass and notch filtering, visual inspection and rejection of noisy channels, artifact removal using \abbrICA{}, and downsampling prior to analysis.

\subsection{Auditory Stimuli}

The speech stimuli from all datasets were processed using a gammatone filterbank implemented in the Eelbrain toolkit \citep{eelbrain}. The filterbank comprised 128 frequency bands spanning 80--\SI{15000}{\hertz} according to the equivalent rectangular bandwidth scale, with a temporal resolution of \SI{1}{\milli\second}. The acoustic speech envelope was computed by summing the absolute value of the gammatone representation across frequency bands. The implementation follows the open-source predictor generator pipeline provided for the Alice dataset \citep{Brodbeck2023Alice}.

\subsection{Final Processing}
\label{sec:finalProcessing}

To ensure consistent analyses across all datasets, an identical final preprocessing pipeline was applied to both the \abbrEEG{} recordings and the acoustic envelope predictors. This consisted of band-pass filtering (1--\SI{8}{\hertz}, using MNE-Python's default finite impulse response (\abbrFIR) filter settings, employing a Hamming window (firwin) and zero-phase (non-causal) filtering), removal of the first and last second of each recording to avoid filter edge effects, downsampling to \SI{50}{\hertz}, and $z$-score normalization.

\subsection{Dataset Categorization}

To facilitate comparisons across datasets, the experimental conditions were grouped according to the primary source of listening difficulty.

\begin{itemize}

\item \textbf{Signal degradation datasets (I, II, VI):}
Listening difficulty was manipulated through changes in the acoustic environment, including \abbrSNR{} and reverberation. These manipulations primarily affect bottom-up auditory processing and auditory scene analysis.

\item \textbf{Scenario dataset (III):}
Listening difficulty was manipulated through changes in task demands, including sustained attention, attentional switching, and audiovisual conversational listening. These manipulations primarily engage top-down attentional and cognitive processes.

\item \textbf{Noise reduction datasets (IV--V):}
Listening conditions differed according to \abbrHA processing, specifically whether {\abbrNR} was enabled or disabled.

\end{itemize}

\section{Methods}
\label{sec:met}

\subsection{Temporal Response Functions}

The temporal response function (\abbrTRF) is a linear finite impulse response (\abbrFIR) model commonly used to model neural speech tracking \citep{13o2015attentional, alickovic2019tutorial, mtrftoolbox}. In auditory attention decoding (\abbrAAD), where the goal is to decode the attended speaker, a backward direction is used to reconstruct the speech envelope $\hat{x}(t)$ from the \abbrEEG signal $y(n,t)$:
\begin{equation}
    \label{eq:backward}
    \hat{x}(t) = \sum_{n} \sum_{\tau} \omega(\tau,n) y(n,t-\tau),
\end{equation}
\noindent
where $n$ denotes the \abbrEEG channel, $t=[1,\ldots,T]$ the time index, and $\omega(\tau,n)$ the decoder weights. The time lags $\tau$ ranged from $-0.6$ to \SI{0.0}{\second}, and the weights were represented as weighted sums of Hamming-windowed basis functions with a width of \SI{50}{\milli\second}, centered on each \abbrTRF lag. These weights were estimated using a sparse boosting algorithm \citep{zhang_boosting_2005,david_estimating_2007} implemented in Eelbrain \citep{eelbrain}.

\subsection{Statistical Analysis}

Model performance was evaluated using cross-validation. Datasets containing up to 12 trials per condition employed leave-one-trial-out cross-validation, whereas datasets with more trials used an approximately 75--25\% train-test split. Performance was quantified using the Pearson correlation coefficient:

\begin{equation}
r_{\mathbf{x}, \mathbf{y}} =
\frac{\sum_i (x(i)-\bar{x})(y(i)-\bar{y})}
{\sqrt{\sum_i (x(i)-\bar{x})^2 \sum_i (y(i)-\bar{y})^2}},
\end{equation}
\noindent
where the reconstructed speech envelope was correlated with the corresponding ground-truth speech envelope.

To investigate whether statistically significant differences exist between the ``easiest'' and ``hardest'' listening conditions, these conditions must first be defined. One approach would be to use participants' subjective difficulty ratings. However, such ratings were not available for all datasets, making a consistent classification across datasets impossible. Instead, we used the mean attended-only reconstruction correlation as a common performance-based measure. Within each dataset, the condition with the lowest mean correlation was hereafter defined as the ``hardest'' condition, while the condition with the highest mean correlation was defined as the ``easiest'' condition.

The effect of listening difficulty was assessed by computing the within-participant difference between the hardest and easiest condition (hardest -- easiest) for each component. These paired differences were evaluated using three complementary statistical measures:

\begin{itemize}
    \item[] \textbf{One-sample $t$-test:} Tested whether the mean difference across participants differed significantly from zero, indicating a systematic increase or decrease between the easiest and hardest conditions.
    \item[] \textbf{Cohen's $d$:} Calculated as the mean difference divided by the standard deviation of the paired differences, providing a standardized measure of effect size.
    \item[] \textbf{Binomial sign test:} Tested whether the proportion of participants exhibiting a positive difference differed from chance (50\%), considering only the direction of change. This served as a non-parametric complement to the $t$-test and is less sensitive to outliers and departures from normality.
\end{itemize}

Together, these measures quantify both the magnitude ($t$-test and Cohen's $d$) and the consistency (sign test) of the observed effects across participants.

\subsection{Attended-Ignored Speech Mixture}

It has long been established that the brain tracks the envelope of attended speech more strongly than that of ignored speech. Using a backward \abbrTRF{} model, this is reflected by higher decoding correlations for the attended than the ignored speech stream \citep{13o2015attentional}. In contrast, little is known about how weighted combinations of attended and ignored speech influence reconstruction performance. Building on the framework proposed by \citet{Wilroth2026IFAC}, we systematically varied the contribution of the ignored speech envelope to the attended speech envelope. For each dataset and condition, the backward decoding correlation was computed using the mixed speech envelope
\begin{equation}
\label{eq:attKign}
    \mathbf{x}_{\mathrm{a+ki}} =
    \mathbf{x}_{\mathrm{att}} +
    k\,\mathbf{x}_{\mathrm{ign}},
\end{equation}
where $k \in [0,1]$ denotes the gain applied to the ignored speech envelope, evaluated in increments of 0.05. As control analyses, decoding correlations were also computed for
\begin{equation}
\begin{split}
\mathbf{x}_{\mathrm{i+ka}} &=
\mathbf{x}_{\mathrm{ign}} +
k\,\mathbf{x}_{\mathrm{att}},\\
\mathbf{x}_{\mathrm{a+kc}} &=
\mathbf{x}_{\mathrm{att}} +
k\,\mathbf{x}_{\mathrm{control}},
\end{split}
\end{equation}
where $\mathbf{x}_{\mathrm{control}}$ denotes the envelope of an unrelated (random) speech signal. The mixed speech envelopes were $z$-score normalized before being used as input to the backward \abbrTRF{} model in \eqref{eq:backward}.

\subsection{Reconstructed Curve Metrics}

\begin{figure}[t]
\centering
\includegraphics[width=0.95\linewidth]{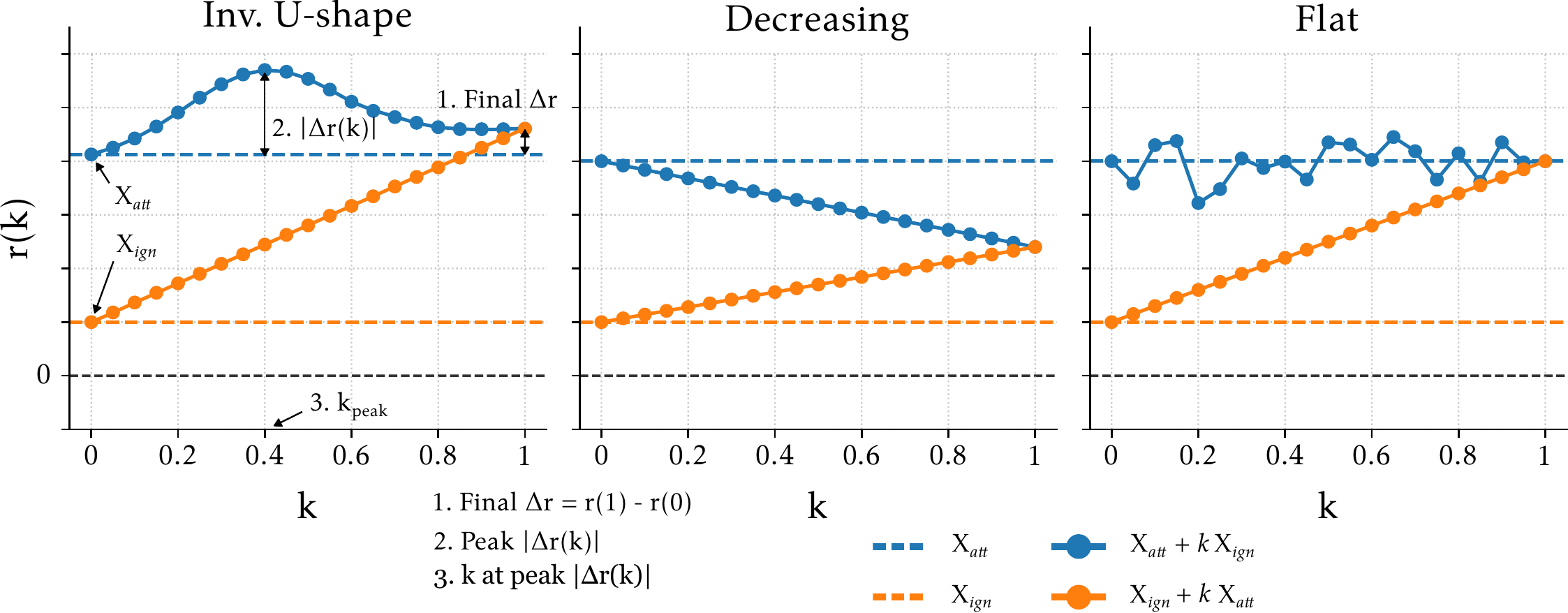}
\caption{\textbf{Illustration of the curve classifications (inverted U-shape, decreasing, and flat), and the three derived components.} The $y$-axis shows the Pearson correlation coefficient ($r$) between the reconstructed and reference speech envelopes, while the $x$-axis shows the ignored-speech gain factor ($k\in[0,1]$). At $k=0$, no speech mixing is applied, corresponding to the attended-only ($X_{\mathrm{att}}$; dashed blue line) and ignored-only ($X_{\mathrm{ign}}$; dashed orange line) reference envelopes. The solid curves show decoding performance obtained from weighted mixtures of the attended and ignored speech envelopes. For inverted U-shaped relationships, Component~1 denotes the change in decoding correlation between $k=1$ and $k=0$, and Components~2--3 quantify the absolute peak improvement, and the optimal gain factor ($k_{\mathrm{peak}}$), respectively.} \label{fig:figure2}
\end{figure}

 The change between a $k$-weighted mixture of attended and ignored speech and the full attended speech ($k=0$) is defined as:

\begin{equation}
    \Delta r(k) = r(k)-r(0).
\end{equation}
\noindent
We identified four characteristics that may reflect differences in acoustic difficulty, as illustrated in Figure~\ref{fig:figure2}.

\begin{enumerate}

    \item \textbf{Final $\Delta r$:} The change in decoding correlation between full attended speech ($k=0$) and equal amplitudes of the attended and ignored speech ($k=1$):
    
    \[
    \text{Final }\Delta r = r(1)-r(0).
    \]

    \item \textbf{Peak $|\Delta r|$:} The maximum absolute increase in decoding correlation relative to the attended baseline,
    \[
       \operatorname*{arg\,max}_{k} |\Delta r(k)|. 
    \]
    Larger values indicate greater sensitivity to the addition of ignored speech.

    \item \textbf{Optimal gain $k_{\mathrm{peak}}$:} The value of $k$ at which the peak correlation occurs. Larger values indicate that higher levels of ignored speech are required to reach the peak.

    \item \textbf{Shape:} The overall shape of the correlation curve was classified as \emph{U-shape}, \emph{inv. U-shape}, \emph{increasing}, \emph{decreasing}, or \emph{flat}.

\end{enumerate}

\subsection{Classification of Reconstructed Curve Shapes}

The classification of the curves was performed by fitting, for each participant,
both a linear model and a second-order polynomial across $k$:

\begin{align}
    r(k) &= \beta_2 k^2 + \beta_1 k + \beta_0, \\
    r(k) &= \alpha_1 k + \alpha_0.
\end{align}

For each condition, the participant-specific quadratic coefficients $\beta_2$ and linear slopes $\alpha_1$ were tested against zero using two-sided one-sample $t$-tests. The sign of the mean coefficient was used to determine the direction of the effect. Curve classification followed a hierarchical procedure in which significant quadratic effects were evaluated before linear effects. 

A curve was classified as inverted U-shaped when the quadratic coefficient satisfied $\beta_2 < 0$ ($p<0.05$), the fitted quadratic vertex was located within the interval $k \in [0.1, 0.9]$, and both endpoints were sufficiently below the peak. Similarly, a curve was classified as U-shaped when $\beta_2 > 0$ ($p<0.05$), the vertex was located within $k \in [0.1, 0.9]$, and both endpoints were sufficiently above the trough. If no significant quadratic effect was detected, significant positive or negative linear slopes were classified as increasing or decreasing, respectively. Non-significant curves were classified as flat. The curve classification criteria is summarized in Table~\ref{tab:curvefitting}.

\begin{table}[t]
\centering
\caption{\textbf{Curve classification.} Significance: $p<0.05$. Vertex inside $[0.1, 0.9]$ ensures the peak is actually interior, not just at an endpoint.}
\label{tab:curvefitting}

\renewcommand{\arraystretch}{1.5}
\begin{tabularx}{0.75\linewidth}{l l}
\toprule
\textbf{Shape} &  \textbf{Criteria} \\
\midrule
Inverted U-shape & $\beta_2<0$, $p<0.05$, vertex inside $[0.1, 0.9]$, endpoints sufficiently below peak \\
U-shape & $\beta_2>0$, $p<0.05$, vertex inside $[0.1, 0.9]$, endpoints sufficiently above trough \\
Increasing & $\alpha_1 > 0$, $p<0.05$\\
Decreasing & $\alpha_1 < 0$, $p<0.05$ \\
Flat & None of the above \\
\bottomrule
\end{tabularx}
\end{table}

\subsection{Partial Correlation Analysis of Ignored Speech} \label{sec:partial_correlation}

For each attended-ignored speech mixture, a separate backward decoder was trained to reconstruct the corresponding target envelope. This approach allowed us to investigate how the relative contribution of attended and ignored speech to the training target influences the learned \abbrEEG-to-envelope mapping. Importantly, because the decoder was retrained for each mixture weight $k$, changes in the reconstructed envelope reflect changes in the learned decoder rather than solely changes in the reference envelope used for evaluation.

A potential complication is that the attended and ignored speech envelopes may themselves be correlated. Consequently, a correlation between the reconstructed envelope and the ignored envelope does not necessarily indicate an association with ignored speech beyond that attributable to the attended envelope. To account for this linear dependence, we employed partial correlation, an approach previously used in neural speech-tracking research to distinguish contributions from correlated stimulus representations \citep{Teoh682,Crosse2021}.

Specifically, we quantified the correlation between the reconstructed speech envelope  $\hat{y}_k$ and the ignored speech envelope $i$ after removing their linear associations with the attended speech envelope $a$, using the partial correlation coefficient:

\begin{equation}
    r_{\hat{y}_k,i \cdot a}
    =
    \frac{
        r_{\hat{y}_k,i}
        -
        r_{\hat{y}_k,a}r_{a,i}
    }{
        \sqrt{
            \left(1-r_{\hat{y}_k,a}^{2}\right)
            \left(1-r_{a,i}^{2}\right)
        }
    }.
    \label{eq:partial_correlation}
\end{equation}

\noindent
The Pearson correlation between the reconstructed and ignored speech envelopes is denoted $r_{\hat{y}_k,i}$, while $r_{\hat{y}_k,a}$ denotes the correlation between the reconstructed and attended speech envelopes, and $r_{a,i}$ refers to the correlation between the attended and ignored speech envelopes. Correlations were calculated using the cross-validated reconstructions, thereby evaluating the learned decoders on data not used for their corresponding model fitting.

We examined the partial correlation for the attended-only decoder ($k=0$). Because this decoder was trained exclusively to reconstruct the attended speech envelope, this analysis assessed whether its reconstruction retained an association with the ignored envelope beyond the linear association attributable to the attended envelope. A significantly positive partial correlation would indicate a positive residual association with ignored speech, whereas a significantly negative partial correlation would indicate a negative residual association. A value not significantly different from zero would provide insufficient evidence for such an association. Importantly, partial correlation removes only the linear dependence associated with the attended envelope at zero lag. It does not necessarily account for nonlinear relationships or temporally shifted dependencies between the speech envelopes. Therefore, a significant partial correlation should not be interpreted as conclusive evidence of an independent neural representation of ignored speech.

For statistical analysis, individual partial-correlation coefficients were transformed using Fisher's $z$-transformation:

\begin{equation}
    z_k
    =
    \operatorname{arctanh}
    \left(r_{\hat{y}_k,i \cdot a}\right)
    =
    \frac{1}{2}
    \ln\left(
        \frac{1+r_{\hat{y}_k,i \cdot a}}
             {1-r_{\hat{y}_k,i \cdot a}}
    \right).
    \label{eq:fisher_z_partial}
\end{equation}

\noindent
The transformation was applied before group-level averaging and statistical testing. Statistical significance was evaluated at the group level, where the Fisher-$z$-transformed partial correlations were tested against zero using two-sided one-sample $t$-tests, with $p$-values adjusted using the Benjamini-Hochberg false-discovery-rate procedure separately for the baseline and training-effect analyses \citep{benjamini1995controlling}. Group-level estimates and their 95\% confidence intervals were transformed back to the correlation scale for visualization.

\section{Results}
\label{sec:res}

In this section, we present the results of our analyses. First, Section~\ref{sec:results_ranking} reports the mean attended-speech decoding correlations across participants, which are used to rank the conditions within each dataset from highest to lowest correlation. This ordering is subsequently used throughout the results figures. Before examining the effects of speech mixing, we evaluate the association between the reconstructed speech envelopes and the ignored speech at $k=0$. Finally, Sections~\ref{sec:results_H1} to \ref{sec:results_H4} present the results corresponding to our three hypotheses (H1--H3) and exploratory question (EQ1).

\subsection{Attended-Speech Tracking Distinguishes Listening Difficulty} \label{sec:results_ranking}

\begin{figure}[t]
\centering
\includegraphics[width=0.95\linewidth]{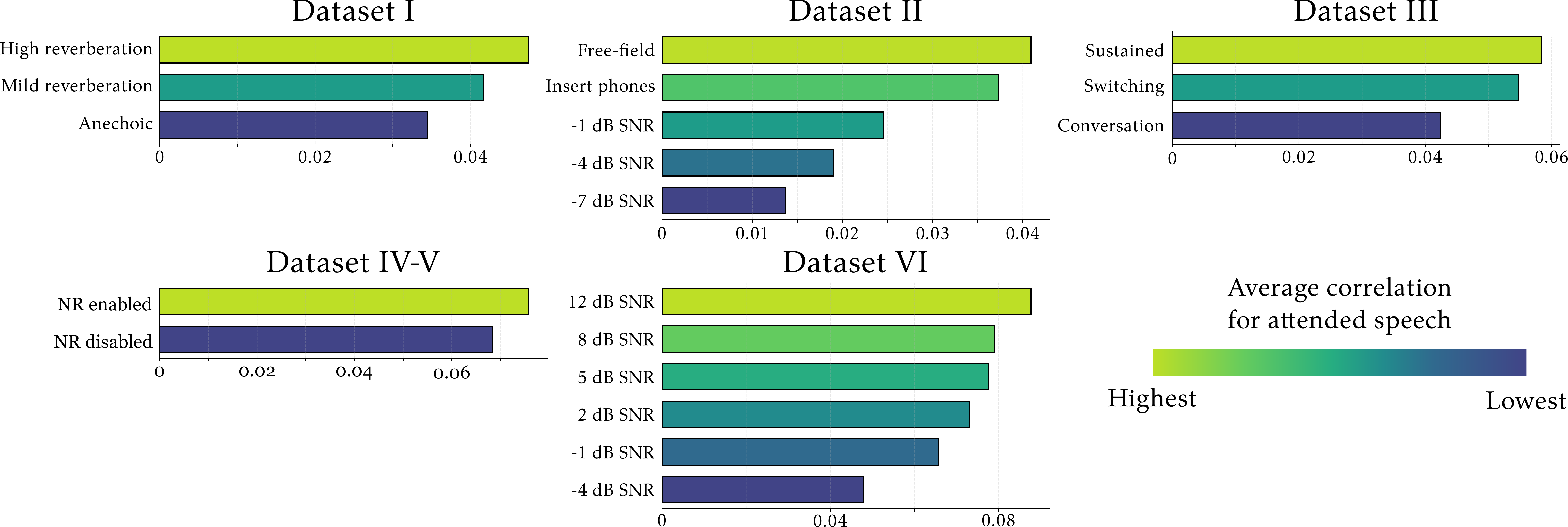}
\caption{\textbf{Attended-speech tracking across listening conditions.} Mean attended-speech correlation across participants for each condition and dataset. Within each dataset, conditions are ordered from the highest (light green) to the lowest (dark blue) attended-speech correlation, providing the within-dataset ranking of listening challenge used in the subsequent analyses of attended-competing speech weighting.}
\label{fig:figure3}
\end{figure}

Attended-speech tracking varied across listening conditions within each dataset (Figure~\ref{fig:figure3}). Conditions were ordered according to the mean attended-speech correlation across participants, from the highest (light green) to the lowest (dark blue) correlation. This ordering provides an empirical index of the relative listening challenge within each dataset and is used throughout the subsequent analyses of the attended–competing speech balance. Because the ordering is based on attended-speech tracking rather than an independent measure of perceived difficulty, comparisons are restricted to conditions within each dataset.

\subsection{Competing-Speech Associations Remain After Controlling for Attended Speech}
\label{sec:partialCorr}

\begin{figure}[t]
\centering
\includegraphics[width=0.95\linewidth]{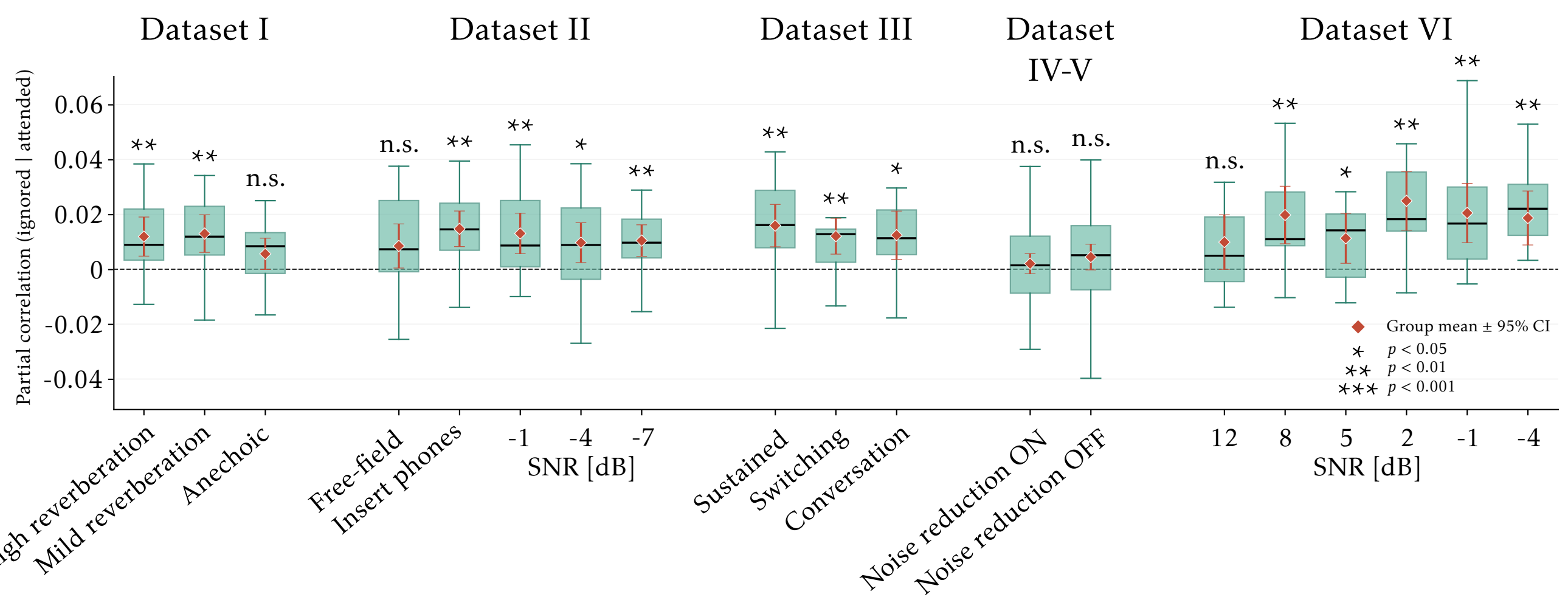}
\caption{\textbf{Ignored-speech association in reconstructed speech envelopes.}
Partial correlations between the reconstructed and ignored speech envelopes were obtained from decoders trained exclusively on the attended speech envelope ($k=0$), while controlling for the attended speech envelope. Boxplots show the participant-level distributions, with boxes representing the interquartile range, black horizontal lines indicating the median, and whiskers extending to 1.5 times the interquartile range. Red diamonds with whiskers indicate group-level estimates with 95\% confidence intervals, calculated on the Fisher-$z$ scale. Group-level estimates and confidence intervals were back-transformed to the correlation scale for visualization. Asterisks indicate statistical significance (*$p<0.05$, **$p<0.01$, ***$p<0.001$) after Benjamini-Hochberg false-discovery-rate correction \citep{benjamini1995controlling}.
}
\label{fig:figure4}
\end{figure}

The results of the partial-correlation analysis are presented in Figure~\ref{fig:figure4}. It shows the partial correlations between the reconstructed and ignored speech envelopes, controlling for the attended speech envelope, for decoders trained exclusively on attended speech ($k=0$). The group-level partial correlations were positive across all investigated conditions, although considerable inter-participant variability was observed. Significant positive associations ($p<0.05$) were found in 14 of the 19 conditions after correction for multiple comparisons, indicating that attended-only reconstructions retained a small association with ignored speech beyond that explained by the linear relationship with the attended envelope. The pooled conditions from Datasets~IV and V exhibited comparatively small partial correlations, neither of which reached statistical significance.

\subsection{\textbf{H1:} The Attended–Competing Balance Vaires Across Listening Conditions} \label{sec:results_H1}

\begin{figure}[t]
\centering
\includegraphics[width=0.95\linewidth]{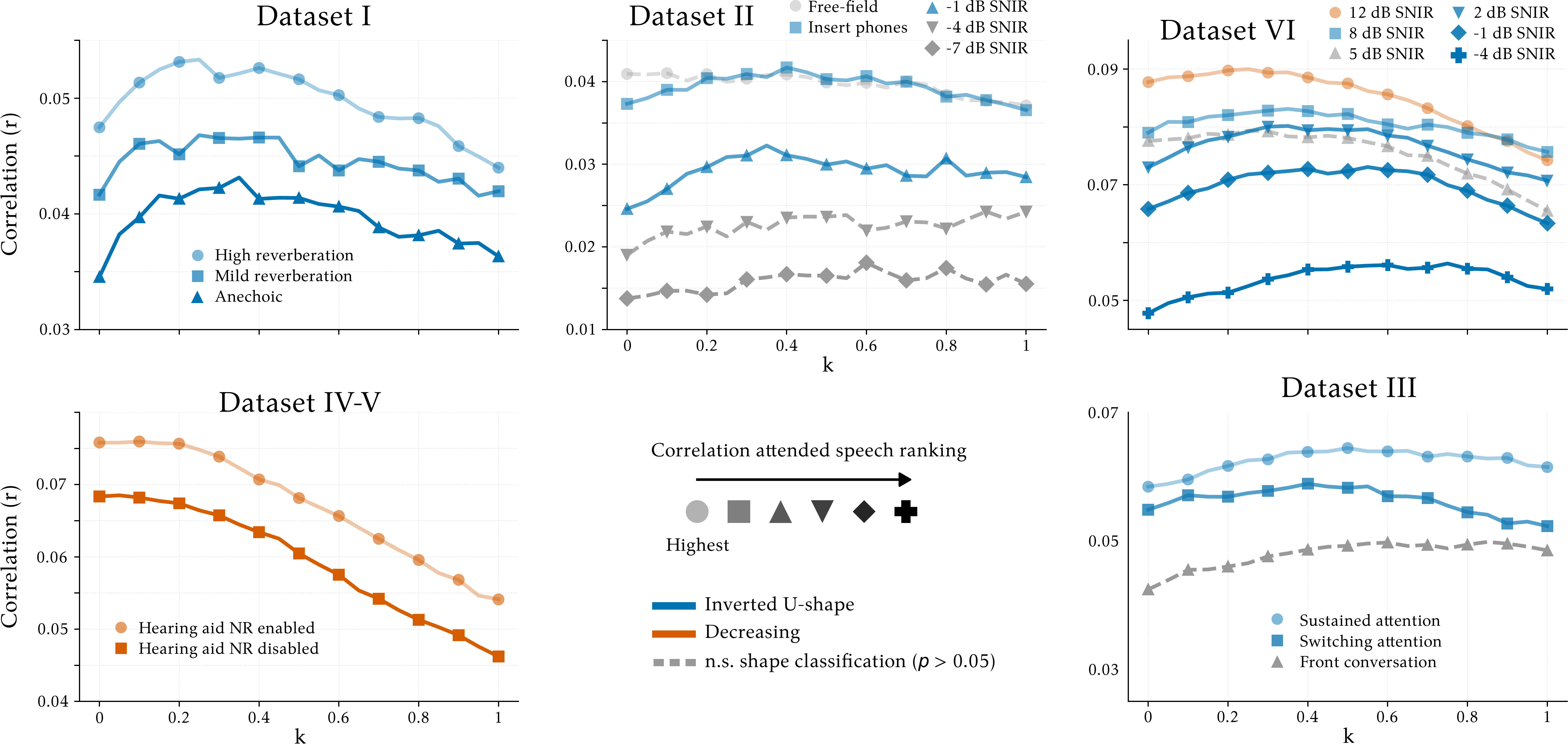}
\caption{\textbf{Attended-competing speech weighting across listening conditions.} Mean cortical speech-tracking correlation as a function of the competing-speech weight $k$ for each condition and dataset. At $k=0$, the weighting contains attended speech alone; at $k=1$, attended and competing speech have equal weights. Solid blue lines denote significant inverted-U relationships, solid orange lines denote significant decreasing relationships, and dashed grey lines denote functions that did not meet the significance criteria for either classification. Within each dataset, conditions follow the ordering established from attended-speech tracking in Figure~\ref{fig:figure3}, with lighter to darker shading indicating increasing listening challenge.}
\label{fig:figure5}
\end{figure}

Cortical speech tracking depended on the relative weighting of attended and competing speech (Figure~\ref{fig:figure5}). Significant ~{inverted-U} relationships were observed across multiple listening conditions. In these conditions, tracking increased when competing speech was added to attended speech, reached a maximum at a non-zero competing-speech weight ($k>0$), and decreased as the contribution of competing speech increased further. Thus, attended speech alone did not consistently provide the strongest tracking; a measurable contribution of competing speech was present in the weighting associated with the maximum. 

The contribution of competing speech associated with maximal tracking was substantial in several conditions, but was not constant across the data. The maxima of the inverted-U functions occurred at different values of $k$, indicating that the relative weighting of attended and competing speech associated with strongest tracking varied across listening conditions. Moreover, the benefit of competing speech was bounded: increasing its contribution beyond the maximum reduced tracking, showing that stronger tracking did not simply result from adding progressively more competing speech. 

Not all listening conditions showed an inverted-U relationship. Significant decreasing functions were also observed, for which tracking was highest at or near $k=0$, and declined as the contribution of competing speech increased. Other conditions did not meet the statistical criterion for either curve classification. The coexistence of inverted-U and decreasing functions therefore indicates substantial variation in how competing speech contributes to tracking across listening conditions.

\subsection{\textbf{H2:} Acoustic and Attentional Manipulations Affect the Attended–Competing Balance Differently} \label{sec:results_H2}

The results for components 1--3 (Figure~\ref{fig:figure2}) are presented in Figure~\ref{fig:figure6}. Each plot shows the mean value for Datasets I--III, and VI, ordered according to the mean attended-only correlations within each dataset as presented in Figure~\ref{fig:figure3}. Due to the decreasing curve classification for Datasets IV--V in Figure~\ref{fig:figure2}, these datasets are excluded from this analysis and are instead evaluated in Sections~\ref{sec:results_H3}~and~\ref{sec:results_H4}.

Statistical comparisons between the easiest and hardest conditions are presented in Figure~\ref{fig:figure7}. To increase the sample size ($n$), data from Datasets I, II, III, and VI, were pooled for each component. Note that the -\SI{7}{\decibel} \abbrSNR condition in Dataset~II was excluded from the statistical analysis due to its very low attended-speech tracking.

Component 1 quantifies the difference in reconstruction performance between the attended-only envelope ($k=0$) and an equal mixture of the attended and ignored speech envelopes ($k=1$). Positive values indicate that incorporating the ignored speech envelope improves reconstruction, whereas negative values indicate that the attended-only envelope provides a better reconstruction. Components 2--3 quantify the absolute peak improvement, and the optimal gain factor ($k_{peak}$), respectively.

\subsubsection{Acoustic Conditions (Datasets I, II, VI)} \label{res:C1}

In this subsection, we present the results for the signal degradation datasets (I, II, VI), based on Figure~\ref{fig:figure6} and Figure~\ref{fig:figure7}.

\begin{itemize}
    \item[] \textbf{Component 1:} The datasets exhibited a significant increase in this measure from the easiest to the hardest conditions (Figure~\ref{fig:figure7}). The correlation difference shifted from negative values under the easiest conditions to positive values under the most challenging conditions, indicating that the benefit of including ignored speech increased with acoustic degradation. In short: As acoustic challenge increases, an equal attended–competing mixture becomes progressively more favorable.
    \item[] \textbf{Component 2:} The datasets have an overall increasing trend from easiest to hardest conditions (significant difference presented in Figure~\ref{fig:figure7}), with the exception of the reportedly very difficult condition of -\SI{7}{\decibel} in Dataset~II. In short: The maximum benefit obtained by including competing speech generally increases with acoustic challenge.
    \item[] \textbf{Component 3:} The datasets exhibit an increasing optimal $k$-value from the easiest to the hardest conditions, indicating that a greater proportion of ignored speech is incorporated into the decoder as task difficulty increases. One possible explanation is that, at lower \abbrSNR values (i.e. under more challenging listening conditions) listeners become less able to suppress the ignored speech. In short:  The weighting associated with maximal tracking generally shifts toward a greater competing-speech contribution.
\end{itemize}

In summary, the results for the signal degradation datasets suggests that increasing acoustic challenge lead to an increased contribution of competing speech.

\subsubsection{Attentional Conditions (Dataset III)}  \label{res:C1}

In this subsection, we present the results for the scenario Dataset~III, based on Figure~\ref{fig:figure6} and Figure~\ref{fig:figure7}.

\begin{itemize}
    \item[] \textbf{Components 1--3:} In contrast to the signal degradation datasets, the pattern was less pronounced for the scenario dataset, where the switching attention condition (ranked \#2) in all three components exhibited lower results compared to sustained attention and conversational attention. 
\end{itemize}

In other words, the attended–competing balance also differed across attentional conditions, but it did not follow the progression observed with increasing acoustic challenge.

\subsection{\textbf{H3:} Acoustic-Condition Dependence is Also Present in Hearing-Impaired Listeners} \label{sec:results_H3}

The hearing impaired participants in Dataset VI wore standard {\abbrHA}s without any {\abbrNR} in conditions with manipulated acoustic challenges. The trend presented in Figure~\ref{fig:figure7} (i.e., all components 1--3 are increasing) suggests that the shift toward a greater competing-speech contribution with increasing acoustic challenge was also observed in hearing-impaired listeners, meaning that the predicted general \abbrHI-specific difference is not evident from these data.

\subsection{\textbf{EQ1:} The Attended–Competing Balance Across {\abbrHA} Processing Conditions} \label{sec:results_H4}

\begin{figure}[t]
\centering
\includegraphics[width=0.95\linewidth]{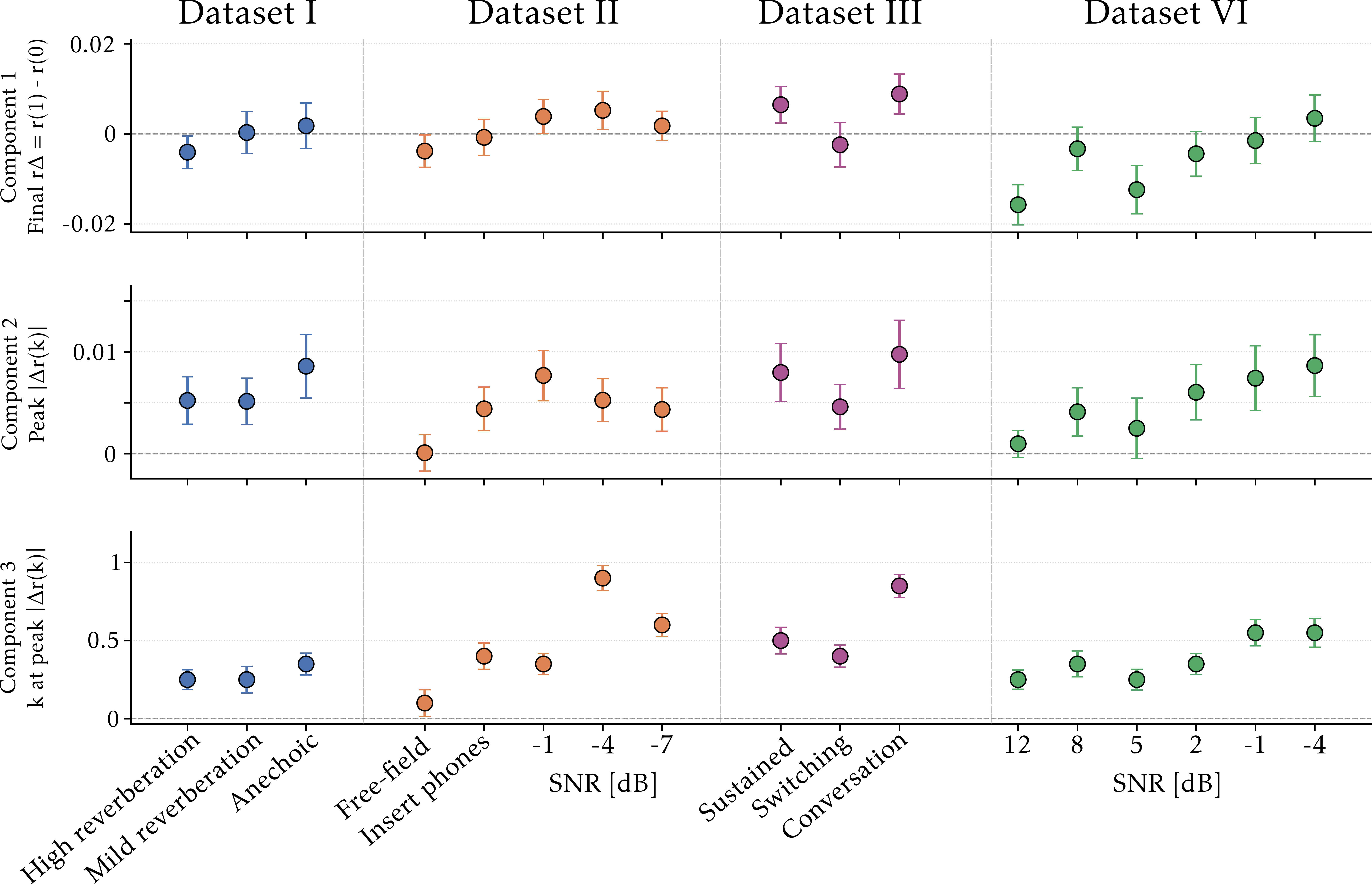}
\caption{\textbf{Results for components 1--3 for each dataset and condition.} Within each dataset, the conditions are ordered from the easiest (left) to the hardest (right). Marker shape and colour indicate the corresponding shape classification shown in Figure~\ref{fig:figure5}.}
\label{fig:figure6}
\end{figure}

\begin{figure}[t]
\centering
\includegraphics[width=0.95\linewidth]{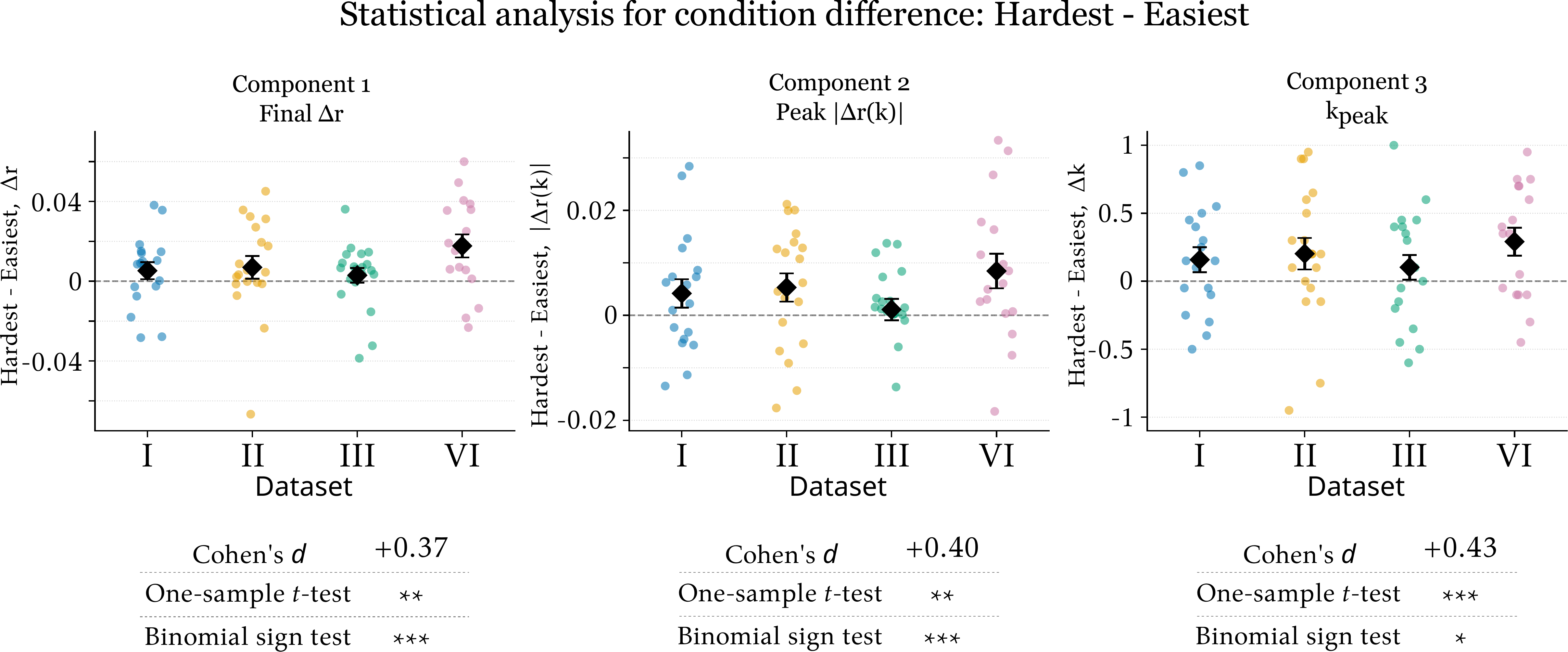}
\caption{\textbf{Statistical analysis of the difference between the hardest and easiest conditions for components 1--3.} To increase the sample size ($n$), the statistical analyses were performed on pooled data for the Datasets I, II, III, and VI. Note: Datasets IV--V (decreasing curve; $n = 65$) showed no significant differences for any component ($p > 0.05$). Asterisks indicate statistical significance after Benjamini-Hochberg false-discovery-rate correction ($N=6$, two tests and three components), where \\ *$p<0.05$, **$p<0.01$, and ***$p<0.001$.
}
\label{fig:figure7}
\end{figure}

Datasets IV--V were collected from hearing-impaired participants wearing {\abbrHA}s with either {\abbrNR} enabled or disabled. The shape was significantly classified as decreasing for both conditions, presented in Figure~\ref{fig:figure6}. This suggests that the attended–competing relationship was similar across {\abbrNR} settings. However, since Dataset~VI also involved hearing-impaired participants, we need to be careful to attribute this decreasing curve classification to hearing impairment, {\abbrHA} use, or {\abbrNR} algorithms alone.

\section{Discussion}
\label{sec:dis}
\subsection{Summary of the Main Findings and Hypotheses}

The central finding is that the attended-competing balance associated with cortical speech tracking is condition dependent rather than fixed. H1 was partially supported: fitted maxima occurred at non-zero competing-speech weights in multiple listening conditions, but decreasing and nonsignificant functions showed that attended speech was not complemented by competing speech in the same way across all conditions. H2 was supported by distinct forms of condition dependence: the weighting metrics showed a broadly ordered progression across the acoustic conditions, whereas the attentional conditions differed without following a comparable monotonic progression. H3 was supported as a generalisation result, with acoustic-condition dependence also observed in hearing-impaired listeners using {\abbrHA}s. For EQ1, the decreasing functions under both noise-reduction settings and the absence of significant differences in the derived metrics provided limited evidence that the attended-competing balance differed across the {\abbrHA} processing conditions examined. Together, these findings show that competing speech remains reflected in cortical speech tracking, but its relative contribution depends on the listening conditions and cannot be described by a common weighting across acoustic, attentional, and {\abbrHA} processing conditions.

\subsection{\textbf{H1:} The Attended-Competing Balance is Condition Dependent} \label{sec:discussion_H1}

The non-zero fitted maxima observed in several conditions indicate that cortical speech tracking is not always characterised by the attended stream alone. This extends the established finding that cortical activity follows both attended and ignored speech, despite stronger tracking of the attended talker \citep{8DingSimon2012a,13o2015attentional}. In this respect, the weighting functions are consistent with a graded selective attention model: competing speech can remain related to cortical activity while being attenuated relative to attended speech \citep{4TREISMAN1964}. 

The variation in curve shape is central to this conclusion. Inverted-U, decreasing, and nonsignificant functions were all observed, indicating that there is no single attended-competing relationship that generalises across listening conditions. We define an inverted-U classification as a significant negative curvature and an interior fitted maximum; we do not directly test whether that maximum is significantly higher than attended-only tracking. The findings therefore support the presence of non-zero fitted maxima in multiple conditions, but not a universal advantage of including competing speech. H1 is partially supported because both the location and form of the weighting functions depend on the listening context.

The weighting is an analytical characteristic of the decoding model rather than a direct measure of how much competing speech is tracked by the brain. It describes the mixed speech target associated with strongest cortical speech tracking. Scalp \abbrEEG{} combines activity from distributed neural sources, and a backward decoder can use any multichannel information that covaries with the target envelope. The weighting should therefore not be interpreted as a neural percentage, gain, or physiological ratio between responses to attended and competing speech. 

This distinction is particularly relevant when considering the cortical hierarchy of selective speech processing. Both streams can influence cortical activity, while selectivity for the attended talker increases at higher-order processing stages \citep{9DingSimon2012b,10mesgarani2012,11Golumbic2013}. A weighting derived from scalp \abbrEEG{} may consequently combine processes with different degrees of attentional selectivity. The present analysis cannot determine whether a non-zero fitted maximum reflects incomplete suppression, concurrent tracking across processing stages, shared acoustic structure, or a combination of these factors. It establishes that the aggregate relationship between cortical activity and the two speech streams changes across conditions. 

The partial-correlation analysis supports this interpretation from a complementary direction. Attended-only reconstructions remained positively associated with the task-defined ignored-speech envelope in 14 of the 19 conditions after controlling for the attended-speech envelope. Competing-speech-related information was therefore present in many reconstructions even when competing speech was not included in the training target. The conditions with significant decreasing weighting functions (\SI{12}{\decibel} \abbrSNR{} in Dataset~VI, and both conditions in Datasets~IV--V, Figure~\ref{fig:figure5}) also lacked significant ignored-speech associations, raising the possibility that the two findings reflect related properties of the data. This correspondence remains descriptive, and nonsignificant partial correlations do not establish the absence of competing-speech-related information. In addition, partial correlation removes only simultaneous linear dependence attributable to the attended envelope. Nonlinear and temporally shifted relationships between the speech streams remain possible, and the analysis does not identify an independent neural source. The weighted-mixture and partial-correlation analyses therefore converge on a specific conclusion: competing-speech-related information remains detectable in cortical speech tracking across many listening conditions, but neither analysis quantifies a separate neural contribution.

These results partially support H1. A non-zero competing-speech contribution was associated with maximal tracking in many listening conditions, demonstrating that attended speech alone was not universally associated with the strongest cortical speech tracking. At the same time, both the magnitude and direction of the effect varied across conditions. The attended-competing balance associated with maximal tracking is therefore not characterized by a single value of $k$.

\subsection{\textbf{H2:} Acoustic and Attentional Manipulations Show Different Patterns of Condition Dependence} \label{sec:discussion_H2}

The clearest systematic change in the attended-competing relationship occurred across the acoustic conditions. The equal-mixture difference, peak change, and observed $k_{\mathrm{peak}}$ (components 1--3, respectively) generally increased as attended-speech tracking (Figure~\ref{fig:figure3}) decreased. Across the explicit \abbrSNR{} manipulations, less favourable acoustic conditions were associated with a more positive equal-mixture difference, a larger peak change, and a greater competing-speech weighting at the observed maximum. Deviations at the most adverse \abbrSNR{}s indicate that this progression was not uniform across the complete acoustic range. 

A useful framework for considering this result is the distinction between auditory-object formation and object selection \citep{6SHINNCUNNINGHAM2008}. Changes in \abbrSNR{} alter the sensory evidence supporting the separation of concurrent speech streams before attention can act on them as distinct objects. The observed shift in weighting may therefore be related to a changing relationship between the streams as the acoustic scene becomes less favourable. Stream segregation was not measured directly, however, and the weighting functions cannot determine whether poorer \abbrSNR{} reduced perceptual separation, altered suppression of competing speech, changed cortical responsiveness, or affected another aspect of speech processing. The acoustic findings establish condition dependence, not its neural mechanism. 

The reverberation experiment places an essential boundary on this interpretation. Its condition ordering was defined by attended-speech tracking and was not monotonic with the physical amount of reverberation. This is consistent with the original observation that attended-speech tracking and auditory attention decoding can remain robust under reverberation \citep{14Fuglsang2017}. The progression observed for this dataset should therefore be understood relative to its empirical attended-speech ordering, not as evidence that increasing reverberation necessarily increases competing-speech weighting. More generally, an acoustic manipulation cannot be assumed to map directly onto a single neural or perceptual scale of difficulty. 

The attentional conditions reorganised the attended-competing relationship differently. Switching attention produced lower values across the weighting metrics than sustained attention and conversation, while conversation produced the highest numerical values. The attentional conditions therefore differed, but they did not lie on a monotonic continuum comparable to the \abbrSNR{} manipulations. 

This non-monotonicity is scientifically informative. Sustaining attention, switching between talkers, and following a conversation impose different structures on selective listening rather than successive levels of a common demand. Sustained attention requires continued selection of one talker, switching requires disengagement from one stream and engagement with another, and conversational listening combines changing relevance with audiovisual and spatial information. Perceptual-load theory \citep{5lavie1995} and the \abbrELU{} framework \citep{12ronnberg2013} both predict that processing demands depend on the listening situation, but neither requires these particular conditions to follow one common ordering. 

The acoustic and attentional findings should therefore not be characterised as opposite effects. The manipulations differed in design, were conducted in separate experiments, and were not crossed within a common paradigm. H2 is supported because the two types of manipulation showed different forms of condition dependence, not because they establish opposing bottom-up and top-down mechanisms. The broader conclusion is that listening challenge is not a unitary variable for cortical speech tracking. Its source and structure matter for the attended-competing balance.

\subsection{\textbf{H3:} Acoustic-Condition Dependence Extends to Hearing-Impaired Listeners} \label{sec:discussion_H3}

The acoustic dependence of the attended-competing balance was also present in hearing-impaired listeners using {\abbrHA}s. A decreasing function at the most favourable \abbrSNR{} gave way to inverted-U functions with non-zero ($k>0$) fitted maxima at several poorer \abbrSNR{}s, accompanied by an overall progression in the weighting metrics. H3 was therefore supported as a generalisation of the acoustic result to this hearing-impaired sample. 

This generalisation matters because it shows that the changing attended-competing relationship is not restricted to young normal-hearing listeners. In aided listeners, as in the other acoustic datasets, the weighting changed with the acoustic condition. The transition from a decreasing function at favourable \abbrSNR{} to non-zero fitted maxima at several adverse \abbrSNR{}s also argues against treating decreasing weighting functions as a general consequence of hearing impairment or {\abbrHA} use. 

The study was not designed to isolate an effect of hearing impairment. The normal-hearing and hearing-impaired groups were not tested using matched stimuli, tasks, acoustic configurations, or {\abbrHA} processing, and the cross-dataset findings do not establish group equivalence. The defensible inference is narrower: acoustic-condition dependence of the attended-competing balance extends to hearing-impaired {\abbrHA} users.

\subsection{\textbf{EQ1:} Hearing Aid Processing Conditions} \label{sec:discussion_H4}

The examined {\abbrHA} processing conditions produced a comparatively uniform attended-competing profile. At the fixed acoustic \abbrSNR{}, the pooled weighting functions decreased as the competing-speech weight increased, both with {\abbrNR} enabled and disabled. The existing comparisons of the derived metrics did not distinguish the processing conditions. For this specific measure of the attended-competing relationship, evidence for differences between the examined settings was therefore limited. 

This result is not evidence that {\abbrNR} has no cortical effect. Previous work has shown that {\abbrHA} processing can alter tracking of attended speech, competing speech, and background noise \citep{alickovic2020front,ELLIITdataset,Andersen2021,WilrothENEURO}. The weighted-mixture framework asks a different question: whether the joint relationship between attended and competing speech changes across processing conditions. Similar decreasing classifications can coexist with differences in stream-specific tracking, absolute correlation, curve magnitude, or other properties not distinguished by the present metrics. 

Acoustic \abbrSNR{}, {\abbrHA} configuration, and noise-reduction setting should also remain conceptually separate. The \abbrSNR{} was fixed, while {\abbrHA} configuration and noise-reduction setting varied within the available designs. Because the configurations were not examined in one complete factorial design, the common decreasing profile cannot be assigned to a single processing factor. EQ1 therefore remains exploratory: the current data identify a shared overall weighting profile but provide limited evidence for differences across the examined {\abbrHA} processing conditions.

\subsection{Implications for Cortical Speech Tracking and Hearing Devices} \label{sec:dis_methods}

The weighted-mixture framework addresses a question that is not resolved by conventional comparisons of attended- and ignored-speech tracking. Separate tracking estimates establish attentional selectivity and support identification of the attended talker. They do not determine how attended and competing speech jointly relate to cortical activity, or whether that relationship remains stable across listening conditions. The present approach characterises this joint, model-derived balance and shows that it is condition dependent. 

This distinction is relevant to {\abbrHA} signal processing. Identifying the attended talker and deciding how the remaining speech streams should be treated are separate problems. Once a target has been identified, a {\abbrHA} must determine whether competing speech should be strongly attenuated, partially preserved, or selectively enhanced relative to the target. A fixed processing rule may not be appropriate if the attended-competing balance changes with the acoustic scene, attentional context, and {\abbrHA} processing conditions. Competing speech may interfere with the attended talker, but it may also provide conversational context or become relevant when attention shifts. 

The weighting estimated here is not a direct prescription for {\abbrHA} gain or attenuation. It describes the relative weighting associated with cortical speech tracking within the present analysis and has not yet been linked to the balance that best supports speech understanding, listening comfort, situational awareness, or attention switching. Nevertheless, the method is a first step towards quantifying a dimension that is largely absent from current neuro-steered processing: not only which talker should be prioritised, but also how access to competing speech should change with the listening context. 

With further methodological development, perceptual validation, and controlled {\abbrHA} studies, the framework could contribute to estimating context-dependent ranges for the relative treatment of attended and competing speech. Such work could help move neuro-steered {\abbrHA}s beyond target identification towards adaptive processing that supports target clarity while maintaining appropriate access to other talkers in the auditory scene.

\subsection{Limitations} \label{sec:dis_limits}

The cross-dataset design provides breadth at the cost of experimental control. The datasets differ in participant characteristics, acoustic presentation, attentional task, {\abbrHA} use, recording system, trial structure, and original preprocessing. A common final analysis pipeline improves consistency but does not make the datasets equivalent. The strongest inferences are therefore within datasets; comparisons across manipulation types and listener groups remain descriptive. 

Relative listening challenge was indexed by mean attended-speech tracking within each dataset rather than by a common behavioural or subjective measure. This ordering supported consistent visualisation and within-dataset comparison, but it does not define a universal difficulty scale. Differences in tracking may reflect acoustic conditions, task demands, data quality, decoder performance, or several of these factors. 

The attended-competing weighting is model derived. The attended- and competing-speech envelopes are statistically dependent, and the partial-correlation analysis removes only their zero-lag linear association. Nonlinear and lagged dependencies remain possible. The fitted quadratic vertex and observed $k_{\mathrm{peak}}$ also quantify different properties: the former contributes to curve classification, whereas the latter is selected from the sampled weights. Neither should be treated as a precisely estimated physiological optimum. 

Curve classification provides a compact description of the weighting functions but depends partly on heuristic criteria. An inverted-U classification does not directly test whether its fitted maximum differs from attended-only tracking, and a nonsignificant difference does not establish equivalence. Curve shape likewise cannot distinguish among auditory-object formation, attentional selection, incomplete suppression, changes in cortical responsiveness, or other mechanisms. 

Finally, the study did not include matched normal-hearing and hearing-impaired groups under identical acoustic and aided conditions. The hearing-impaired data support generalisation of acoustic-condition dependence, but they do not establish whether hearing impairment modifies that dependence.

\subsection{Future Work} \label{sec:futureWork}

The central next step is to establish the perceptual meaning of the attended-competing weighting. Relating it to speech understanding, comprehension, attention-switching performance, listening effort, and perceived difficulty would determine whether changes in the weighting correspond to meaningful differences in listening or primarily describe properties of the decoding model. 

A common experimental design that crosses acoustic and attentional manipulations would address the distinction raised by H2 directly. Manipulating \abbrSNR{} and attentional task within the same listeners and speech materials would permit evaluation of their separate and interacting effects, while behavioural and subjective measures would provide an independent account of listening challenge. 

Temporally resolved and source-resolved analyses could determine whether the weighting changes across response latency and cortical region. Such analyses would test whether the condition dependence observed at the scalp level is distributed across the response or differs between earlier and later stages of speech processing.

For {\abbrHA} processing, controlled within-participant studies are needed to separate acoustic input, {\abbrHA} configuration, and noise-reduction setting. These studies should establish whether changes in the weighting are related to listener benefit and whether the relevant balance changes when attention moves between talkers. Only after this validation would it be appropriate to examine whether the framework can inform {\abbrHA} evaluation, fitting, or adaptive processing.

\section{Conclusion}
\label{sec:con}
The attended-competing relationship in cortical speech tracking was neither absent nor fixed. Non-zero competing-speech weighting occurred in several conditions, but decreasing and nonsignificant functions showed that this was not a universal feature. Acoustic conditions produced a broadly ordered change in the weighting metrics, attentional conditions reorganised the balance without following the same progression, and the acoustic dependence extended to hearing-impaired listeners using {\abbrHA}s. The {\abbrHA} processing comparisons provided limited evidence for differences across the examined settings. 

The weighting is a model-derived description of the joint relationship between attended and competing speech, not a physiological ratio. Within that boundary, the findings move cortical speech tracking beyond a binary attended-versus-ignored account. Competing speech remains variably related to cortical activity during selective listening, and that relationship changes with the listening context. With further perceptual and {\abbrHA} validation, quantifying this balance may help address not only which talker should be prioritised, but also how access to competing speech should be managed in complex auditory scenes.

\section{Acknowledgements}
\label{sec:ack}

We thank the research audiologists and participants at Eriksholm Research Centre (ERH) for their assistance with data collection. We specifically want to thank Payam Shahsavari (ERH) Baboukani and Oskar Keding (Lund University) for valuable discussions

\section{Declaration of generative AI and AI-assisted technologies in the writing process}

During the preparation of this work the authors used AI in order to improve readability by rephrasing sentences, finding synonyms, and checking grammar and spelling. After using this tool/service, the authors reviewed and edited the content as needed and  take full responsibility for the content of the publication.

\section*{Appendix}
\label{sec:app}

\begin{figure}[h]
\centering
\includegraphics[width=0.95\linewidth]{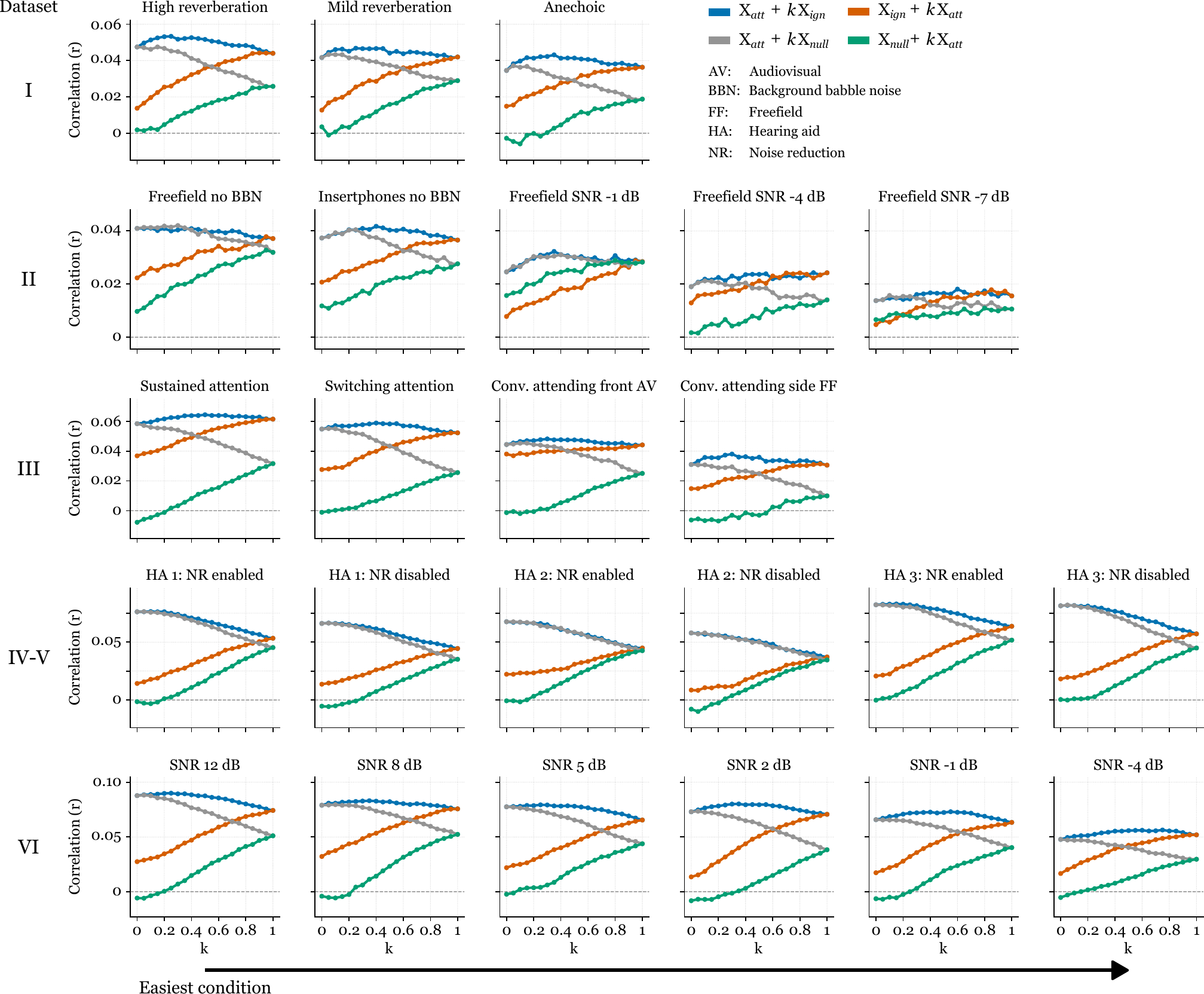}
\caption{\textbf{Average decoding correlations for all speech-mixture combinations.} Rows correspond to datasets and columns to experimental conditions, ordered from the easiest to the hardest according to the results presented in Figure~\ref{fig:figure3}. Blue curves show mixtures of the attended and ignored speech envelopes (also shown in Figure~\ref{fig:figure5}), orange curves show mixtures of the ignored and attended speech envelopes, and the grey and green curves show the corresponding control mixtures with a random (\emph{null}) speech envelope.}
\label{fig:figure8}
\end{figure}

The average decoding correlations for all speech-mixture combinations are presented in Figure~\ref{fig:figure8}.

\newpage

\bibliographystyle{apalike}
\bibliography{refs}

\end{document}